# CREATING A SYNTHESIZER FROM SCHRÖDINGER'S EQUATION

*Arthur Freye, Jannis Müller*

Dep. of Mathematics, Natural Sciences and Informatics
Technische Hochschule Mittelhessen
University of Applied Sciences
35390 Gießen, Germany
**arthur.freye|jannis.mueller@mni.thm.de**

## ABSTRACT

Our project offers an alternative approach to the sensory perception of the Schrödinger equation (an elementary model of quantum phenomena) by interpreting it as a sound wave. We are building a synthesizer plugin that simulates a quantum mechanical state that evolves over time. Thus, our tool allows the creation of unique sounds that are in motion and feel alive. These can be used in professional music production without any knowledge of physics, while at the same time providing insight into a chapter of quantum mechanics. The goal is to lower the threshold for entering complex theory by first developing an intuition for the subject; but the tool can also be used purely as a musical instrument. The user is encouraged, but not forced, to learn more about the underlying physics. Simulation parameters are adjustable in real-time, allowing intuitive experimentation. Despite the approximate calculations, real physical effects such as quantum tunneling can be observed acoustically and visually.

## 1. INTRODUCTION

The subject of quantum mechanics is so vast today that choosing to learn about it means taking on an overwhelming amount of material. Curiosity can quickly turn to resignation. The literature often assumes that students of quantum physics already know a lot about classical physics. Many concepts can only be expressed mathematically and not in words. This makes it difficult to communicate to the general public.

Our goal is to build a tool that allows everyone, even people without a background in physics, to gain insight into quantum mechanical behavior. A sensory approach, in this case through the ear, opens up different ways of understanding as well as possibilities for medial processing and application. The tool should be usable to produce interesting and aesthetic sounds with simple but versatile controls. At the same time, we want to spark the user's interest; they should be intrinsically motivated to learn more about the functionality and concepts behind the sounds they create. Since we are not physicists, but rather driven by curiosity, we are part of our own target audience.

The combination of quantum physics and sound has been explored in other works. An elaboration on sonification choices for simulated data is presented in [1]. This thesis involves complex quantum mechanical models. Our approach is more basic, exchanging provable correctness for simpler explanation and understanding. This allows us to build a practical application. A collection of different approaches can be found in [2] with a focus on quantum computing algorithms that help in music composition and production. Our project is not about quantum computing, but about the model that describes the nature of quanta. We have a similar setting to the work of [3], where qubit measurements are synthesized into music. Instead of measuring qubits, we use Schrödinger's equation, simulate it and present a distributable software tool with flexible parameters. The Schrödinger equation is also used in [4] to explore the connection between quantum states and sound information in both directions. Our project is purely about generating sound from quantum behavior. This approach is also present in [5], which follows a functional and aesthetic goal. We share most of this goal, but we also include a didactic aspect in our project. Furthermore, the two projects use very different sonification decisions for Schrödinger's equation, which can be interesting to compare.

The Schrödinger equation lies at the foundation of quantum mechanics. The subject connects to wave functions, which are also prominent in classical physics, and every person has a basic understanding of simple waves since they are observable in our everyday lives. This motivated us to work with Schrödinger's equation and combine it with sound, which most people have an intuitive access to as well. In the following chapters, we will present how the simulation works and our sonification decisions, how it is implemented, and the results of testing the software. Finally, we give an informal assessment of possible applications and what features could be added, improved, or changed.

## 2. BASICS

Music and Schrödinger's equation also share several qualities from a technical perspective, not just the human perspective mentioned above. To implement it in software, we need to prepare an algorithm that can be computed by a machine and then map it to audio signal parameters. The equation works in the space of complex numbers, but all our inputs (initial wave functions) and values used for sonification (probability density) are real. Complex numbers are not hard to understand, but they are difficult to comprehend as an effect in the Schrödinger equation.

### 2.1. Discretized Schrödinger equation

Schrödinger's equation has multiple forms for different dimensions and settings. We use the time-dependent, one-dimensional version (1), which describes the time evolution of a complex wave function $\psi(x,t)$ (or $\psi$ for short).

$$i\frac{\partial \psi}{\partial t} = V(x)\psi - \frac{\partial^2}{\partial x^2}\psi \tag{1}$$

In (1), $\psi$ is a solution of the differential equation. The potential energy $V(x)$ and kinetic energy $\frac{-\partial^2}{\partial x^2}$ operators are derived from the Hamiltonian operator [6]. $\psi$ can be physically interpreted as the probability amplitude. The squared absolute $|\psi|^2$ represents the probability density. Scaling factors such as mass and the Planck constant are omitted in (1) because they only affect scaling, and we focus on the behavior independent of absolute values.

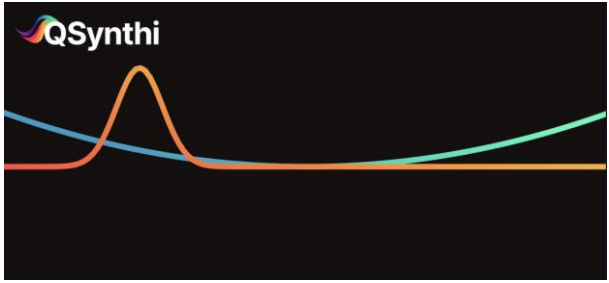

Figure 1: Possible initial state of the wave function $|\psi(x,0)|^2$ (orange), a normal distribution inside a quadratic potential $V(x)$ (blue/green).

To implement (1) into computer algorithms, it must be transformed into a discrete approximation that takes a wave function as input, performs a "timestep" that evolves the function by a given amount of time and outputs the result. There is such an approximation proposed in [7]. It introduces $\Delta t$ as the timestep quantity and splits the time evolution into two parts (2) and (3), one for each energy operator.

$$\widetilde{\Psi}_{t+\Delta t} = \exp(-i\, V(x)\, \Delta t)\, \Psi_t \qquad (2)$$

$$\Psi_{t+\Delta t} = \text{iFFT}\left(\exp\left(-i\left(2\pi\frac{j}{n}\right)^2 \Delta t\right) \text{FFT}(\widetilde{\Psi}_{t+\Delta t})\right) \qquad (3)$$

Equations (2) and (3) now work with a discrete array $\Psi_t$ defined as $\Psi_t(x_j) = \psi(x_j, t)$, $j = 1, \ldots, n$. The array $\widetilde{\Psi}_{t+\Delta t}$ is the state after the potential energy part of the time evolution. A fast Fourier transform (FFT) and an inverse fast Fourier transform (iFFT) are required to convert $\widetilde{\Psi}_{t+\Delta t}$ from the position space to the momentum space and later back again to obtain the desired output [8]. All of these algorithms are discrete and can be implemented in code.

### 2.2. Sonification

The sonification method generates the sound with a given frequency, e.g., determined by a certain key on a keyboard. Unlike some other Schrödinger's equation sonifications, this approach generates the sound from the simulation alone, without using a sound file as input.

The main point is to reinterpret the current state of the simulation, $\Psi_t$, as a single period of a tone, where the x-value represents the phase and $|\Psi_t|^2$ represents the amplitude. This is possible because the simulation only covers a limited domain for $x$. The single period can then be looped at a given frequency to obtain a tone.

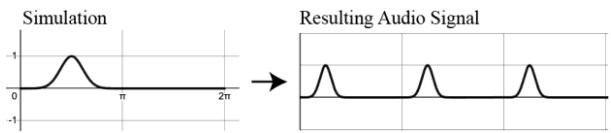

Figure 2: A tone is created by looping a period several times.

While the tone is playing, the simulation runs in the background and updates the state. The new $\Psi_{t+\Delta t}$ becomes the source of the audio output, constantly influencing the shape of the resulting sound wave, changing its characteristics, and ultimately creating an evolving sound.

To make it polyphonic, each time a new note is triggered, a new separate simulation is started based on the current settings for the initial state. The outgoing sound signals could then be amplified with a classic ADSR (Attack, Decay, Sustain, Release) envelope for each voice played and summed to create a single output signal. This method provides a framework that can be applied to any one-dimensional time-based simulation in a limited domain for $x$, such as water or sound waves in a closed environment.

### 2.3. Stereo audio

To preserve as much information as possible through the sonification, an additional step can be taken. Using only the method described in chapter 2.2, all the information about the position of a pattern on the x-axis is lost.

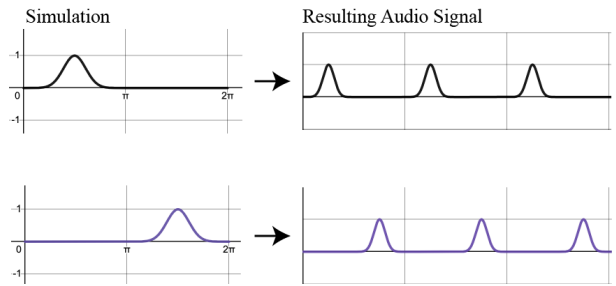

Figure 3: Two simulation states and their resulting audio signals.

As can be seen in Figure 3, two different states result in almost the same sound wave, just out of phase. When played back sequentially, the two signals would sound the same, even though they have different simulation states. To preserve the positional values of the signal, a stereo output can be used, with simulation states that have higher values for $|\Psi_t|^2$ on one side sounding more tilted to that side. In [9, *Harmonic_Oscillator_Demo.mp4*], this effect can be heard with devices that can play stereo audio. When the peak of the orange curve representing $|\Psi_t|^2$ is on the one side, the audio signal on that side is louder. We will discuss two ways to achieve this effect, each with advantages and disadvantages:

The first method is to volume-pan the signal based on the concentration of $|\Psi_t|^2$. This is done by calculating the average of each side and amplifying the signal on the corresponding audio channel. The advantage of this technique is that it introduces almost no distortion into the audio signal. A disadvantage is that it introduces large volume differences between the left and the right audio channels. Additionally, since this method still sends the same signal to both channels, the different characteristics of each side cannot be fully represented in the audio output. In Figure 4, the simulation state is much sharper on the right than on the left. This would be inaudible with volume panning.

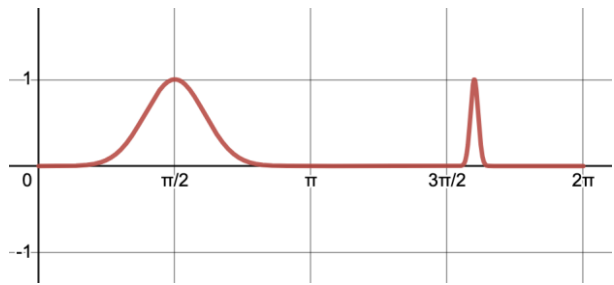

Figure 4: Simulation state where different characteristics on each side are desired for sonification.

The second method takes a different approach. Instead of panning the entire signal, it multiplies the simulation state by a sigmoid-like function for each channel. For the left channel, the function should be 1 at the leftmost point of the simulation and 0 at the rightmost point, and vice versa for the right channel. In the example, a transformed version of the *smootherstep* function $f(x) = 6x^5 - 15x^4 + 10x^3$ is used: $f_r(x) = f\left(\frac{x}{2\pi}\right)$ for the right channel and $f_l(x) = 1 - f\left(\frac{x}{2\pi}\right)$ for the left channel.

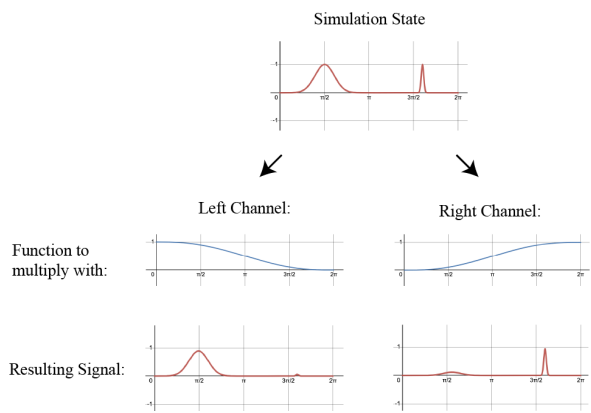

Figure 5: Stereo calculation process with the *smootherstep* function.

As shown in Figure 5, the left part of the simulation is more present in the left channel and vice versa, so the character of each side of the simulation can be preserved. There is a downside to this method, as it introduces audible distortion to wider patterns, such as a single phase of a sine wave spread over the entire simulation range.

The effect can be heard in action in [9, *Tunnel_Effect_Demo.mp4*], where the tunnel effect is sonified. The tunnel effect describes the phenomenon that a quantum object (e.g., an electron) can pass through a high potential energy barrier due to its wave nature; something that is impossible in classical physics. In the example, the accuracy is significantly reduced to enhance the tunneling, which causes some visible errors later in the simulation.

We decided to use the second method in our implementation because it has the advantage of being able to reproduce different characteristics on each audio channel. As long as $f_l(x)$ and $f_r(x)$ are chosen so that $f_l(x) + f_r(x) = 1$ in the simulation range, the signal can be easily converted back to mono without any distortion.

## 3. IMPLEMENTATION

We use C++ with the *juce* framework to build a synthesizer plugin for digital audio workstations (DAW). The framework guarantees seamless communication between our code and any DAW and generates production-ready VST3/AU plugins. There are no additional requirements to integrate and use our tool other than a basic knowledge of the specific music software, which can be easily obtained from free online sources. User settings are versatile and adjustable in real-time. This is designed to increase user engagement and create memorable learning effects.

The main task of the implementation is to provide the DAW with a continuous stream of audio samples based on a MIDI input; MIDI refers to a technical standard for digital communication with electronic musical instruments, devices and computers. When a key is pressed, a simulation is started and sonified at the frequency of the note played, using the method from chapter 2.2. The n-element array $|\Psi_t|^2$ is therefore treated similarly to a wavetable in a wavetable synthesizer. With the note frequency $f_n$ from a MIDI input and the sampling frequency $f_s$ from the DAW, $\frac{f_s}{f_n}$ determines the number of samples in a single period. Since the array $|\Psi_t|^2$ is interpreted as one period, we know that $\frac{1}{f_n}$ evenly spaced values from $|\Psi_t|^2$ are needed for $\frac{1}{f_n}$ seconds of audio. Then $|\Psi_t|^2$ is computed for the required values and linear interpolation is used to retrieve a sample between two discrete array values; this is necessary to sample $|\Psi_t|^2$ for any $f_n$ and $f_s$. Using the C++ modulo operator, the array access can be looped for continuous sampling.

Besides the audio output, our tool also visually displays the wave and potential functions. This helps the user to see how different input parameters affect the simulation. Additionally, the auditory experience can be compared to the visuals to determine where certain sound behaviors originate in the state of the wave.

Several user parameters control the accuracy and speed of the time evolution simulation. For the Fourier transform, we added an FFT library [10] with advanced algorithms for good performance, since two FFTs are computed for each timestep. Within the timestep calculation, the code does not check given inputs for analytical correctness, i.e., whether they are actual solutions to Schrödinger's equation. This allows a wide variety of parameter combinations to be used in the simulation. In most cases, arbitrary inputs don't solve Schrödinger's equation, so no physical results or accuracy can be expected. But with this freedom to experiment, the behavior of the wave evolution can be observed in many scenarios, not just certain restricted ones.

## 4. FINDINGS

As described in chapter 2.2, the sonification process uses only the squared $\Psi_t$ values (state of the wave). The resulting sound wave is therefore deterministic, since no measurements are taken that would lead to the collapse of the wave function. This makes it feasible as a musical instrument because the performer knows what sound to expect when a key is pressed. If a degree of randomness is desired, the digital audio workstation can be used to automate various parameters.

By selecting the initial state of the wave, the characteristics of the sound can be defined. The time evolution is then influenced by parameters such as simulation speed and potential settings. This provides a great deal of flexibility for sounds. Combined with modules familiar from other synthesizers, such as filters and envelopes, this synthesis technique allows the creation of constantly evolving sounds. This makes it possible to create unique and desirable sounds in many situations that would otherwise be unattainable or only possible with very complex modulation. The supplementary material includes a sample video showing the process of designing a sound from scratch [9, *Violin_Demo.mp4*] and a visualization of different sounds [9, *Pluck_Demo.mp4*, *Pad_Demo.mp4*]. Please note that the plugin is still a work in progress.

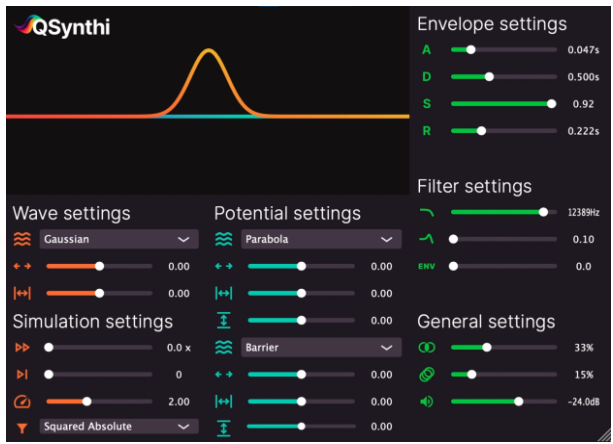

Figure 6: User interface of our synthesizer plugin, with input parameters and visual output (top left).

Interestingly, almost all simulations have patterns that repeat over time, matching the rhythmic nature of music. This can be used in a variety of musical ways. For example, the speed of the simulation can be adjusted so that the pattern repeats in sync with the tempo of a song. It's also possible to go even faster, creating an effect like pulse width modulation, but with much more depth. Alternatively, if the simulation speed is set very low, different notes can be timed so that the evolving patterns are out-of-phase, creating interesting interactions. An example of this effect can be heard in [9, *Counting_Gaussians.mp3*].

Working with simulations based on physical model systems, such as the quantum harmonic oscillator or a quantum tunneling setup, gives results very close to the expected behavior. With the freedom to adjust many options from there, a basic intuition for these models and the behavior of quantum objects can be acquired. This is not definitive knowledge, but rather a simplified view of a difficult subject that can be expanded and refined with further study.

## 5. CONCLUSION

The result of this project is a musical instrument that has the potential to produce aesthetic and interesting sounds, while also providing a simplified and sensory access to basic quantum mechanical behavior. It can be used as a synthesizer in music production and allows to experiment with "quantum-based" sound in a unique way. Since it simulates natural behavior, the question arises whether it actually plays natural sounds instead of synthesized sounds.

With further development of the tool, many additional features are possible. There may be other sonification decisions that produce different results. More musical options such as filters could be added, but it is important to keep the original sound characteristics; smoothing the audio too much goes against our intention to sonify quantum behavior. Simulations of different physical equations can be added, e.g., classical water waves. Listening to the real or imaginary part of the complex number output, as an alternative to the squared absolute, could lead to interesting results as well. The balance of input parameters between intuitive and scientific is difficult and can be tested and adjusted in the future.

We can't teach quantum mechanics with our tool, but an understanding of this simple model can be a motivation to keep exploring and learning. Our implementation shows a feasible way to make a sonified interpretation of Schrödinger's equation and should inspire other sound experiments with this simulation. Since the programming approach is less complex than the mathematics, it can serve as an entry point for others; comparing alternative simulations by sound could also lead to code improvements. Our tool can be found at qsynthi.com.

## 6. ACKNOWLEDGMENT

We thank the faculty for Mathematics, Natural Sciences and Informatics (MNI) of our university, Technische Hochschule Mittelhessen, and especially the dean of the faculty, Prof. Dr. Peter Kneisel. Their generous support made our participation at ICAD 2023 possible. Additional thanks go to Prof. Dr. Dominikus Herzberg for guiding us through the writing process.